\documentclass[journal,12pt,onecolumn,letterpaper]{IEEEtran}
\usepackage{arxiv}
\usepackage{geometry}
\usepackage{times}
\usepackage{cite}
\usepackage{url}
\usepackage{graphicx}
\graphicspath{{Images_SimplyMime/}}
\usepackage{lscape}
\usepackage{subfigure}
\usepackage{rotating}
\usepackage{rotfloat}
\usepackage{xcolor}
\usepackage{amsmath}
\usepackage{amssymb}
\usepackage[linesnumbered,ruled,vlined]{algorithm2e}
\usepackage{pseudocode}
\usepackage{array}
\usepackage[english]{babel}
\usepackage{gensymb}
\usepackage{textcomp}
\usepackage{placeins}
\usepackage{balance}
\usepackage{booktabs}

\usepackage{multirow}

\title{Acoustic Levitation for Environmental Remediation: An Effective Approach for Containment and Forecasting of Oil Spills}

\author{%

L Rochit \\
    Centre of Excellence, Artificial Intelligence \& Robotics (AIR),\\
    School of Computer Science and Engineering\\
    VIT-AP University, India \\
    \texttt{rochitl72@gmail.com}
\And

Nithish Kumar N\\
    Centre of Excellence, Artificial Intelligence \& Robotics (AIR),\\
    School of Computer Science and Engineering\\
    VIT-AP University, India \\
    \texttt{nithishkumar.n1923@gmail.com}
\And
    Devi Priya V S \\
    Department of Computer Science and Engineering\\
    Siddhartha Academy of Higher Education, India \\	
   \texttt{vsdevipriya@gmail.com}
\And

  Sibi Chakkaravarthy Sethuraman\\
    Centre of Excellence, Artificial Intelligence \& Robotics (AIR),\\
    School of Computer Science and Engineering\\
    VIT-AP University, India \\
    \texttt{sb.sibi@gmail.com} \\
\And
  Anitha Subramanian\\
    Centre of Excellence, Artificial Intelligence \& Robotics (AIR),\\
    School of Computer Science and Engineering\\
    VIT-AP University, India \\
    \texttt{anithachubbu@gmail.com} \\
    
}
\begin{document}


\maketitle




\begin{abstract}
 The ocean ecology is badly impacted by large-scale oil spills, plastic waste, and chemical pollution, which destroy ecosystems and endanger marine life. Acknowledging the detrimental effects of oil spills on ecosystems, our research aims to establish the foundation for creative methods to lessen their impact. With an emphasis on the containment and prediction of oil spills, this research investigates the potential of acoustic levitation as a cutting-edge technique for environmental cleanup. Effectively separating and eliminating pollutants without causing additional ecological harm is a major issue for traditional oil spill cleanup techniques. Acoustic levitation provides a non-invasive, accurate, and effective alternative by using sound waves to precisely and subtly separate oil droplets from water in controlled environments. This proposed approach can reduce the negative effects on the environment and increase the efficacy of cleanup efforts. The findings have been examined and assessed by proof of concept experiments with oil droplets, identifying the relationship between the intensity of ultrasonic pressure and the proportion of oil droplets collected. 
\end{abstract}





\section{Introduction}

Oceans are among earth's most valuable natural resources, making up approximately 70 \% of the planet's surface. They control the climate, purify the air, and aid global food production. In 2023, there were ten oil spills totaling more than seven tonnes, according to the ITOPF's Oil Tanker Spill Statistics \cite{news_2024}. A decade average of 6.8 is reached with these statistics. In addition to pipeline spills and oil industry activities, petroleum usage (including oil spills from non-tankers and "run-off" from roads and other land-based sources) and natural seepage all impact annual intakes. Numerous components of socio-ecological systems are impacted by oil contamination \cite{de2021immediate}. By contaminating and bioaccumulating ecological systems, they have the potential to adversely impact ecosystem functioning, the structure of ecological assemblages, and the physiological health of impacted creatures. Oil spills, whether in marine environments, nearshore locations, or ports, demand comprehensive strategies for effective mitigation. The cost of cleanup operations, a primary economic factor, underscores the urgency for ongoing research. The financial impact of oil spill cleanup is profound with nearshore, and port spills, often incurring significantly higher expenses compared to offshore incidents. Understanding the complexities of cleanup operations, especially in areas with a higher probability of shoreline impact, is essential for developing cost-effective and efficient response plans. The economic consequences extend beyond direct cleanup expenses, encompassing broader impacts on industries, ecosystems, and local communities. 

\begin{table*}[h]
\centering
\caption{Cleanup cost of types of oils for spills and  Per-Unit Marine Non-US Oil Spill Cleanup Costs By Primary Cleanup Methodology}
\label{Cleanup cost of types of oils for spills}
\begin{tabular}{|l|c|clcc}
\cline{1-2} \cline{4-6}
\multicolumn{1}{|c|}{\textbf{Oil Type}}                     & \textbf{\begin{tabular}[c]{@{}c@{}}Cost (2009\\  \$ US/t)\end{tabular}} & \multicolumn{1}{c|}{\textbf{}} & \multicolumn{1}{c|}{\textbf{\begin{tabular}[c]{@{}c@{}}Primary \\ Method\end{tabular}}} & \multicolumn{1}{c|}{\textbf{\begin{tabular}[c]{@{}c@{}}US \$/\\ tonne\end{tabular}}} & \multicolumn{1}{c|}{\textbf{\begin{tabular}[c]{@{}c@{}}US \$/\\ Liter\end{tabular}}} \\ \cline{1-2} \cline{4-6} 
\begin{tabular}[c]{@{}l@{}}No.2 Diesel \\ Fuel\end{tabular} & 2940                                                                    & \multicolumn{1}{c|}{}          & \multicolumn{1}{l|}{Manual}                                                             & \multicolumn{1}{c|}{23403.45}                                                        & \multicolumn{1}{c|}{21.03}                                                           \\ \cline{1-2} \cline{4-6} 
Light Crude                                                 & 5435                                                                    & \multicolumn{1}{c|}{}          & \multicolumn{1}{l|}{Dispersions}                                                        & \multicolumn{1}{c|}{5633.78}                                                         & \multicolumn{1}{c|}{5.06}                                                            \\ \cline{1-2} \cline{4-6} 
No.4 Fuel                                                   & 30440                                                                   & \multicolumn{1}{c|}{}          & \multicolumn{1}{l|}{Mechanical}                                                         & \multicolumn{1}{c|}{9611.97}                                                         & \multicolumn{1}{c|}{8.64}                                                            \\ \cline{1-2} \cline{4-6} 
No.5 Fuel                                                   & 29544                                                                   & \multicolumn{1}{c|}{}          & \multicolumn{1}{l|}{\begin{tabular}[c]{@{}l@{}}In Situ \\ Burning\end{tabular}}         & \multicolumn{1}{c|}{3127.87}                                                         & \multicolumn{1}{c|}{2.81}                                                            \\ \cline{1-2} \cline{4-6} 
Crude                                                       & 9236                                                                    & \multicolumn{1}{c|}{}          & \multicolumn{1}{l|}{Natural}                                                            & \multicolumn{1}{c|}{1286.00}                                                         & \multicolumn{1}{c|}{1.15}                                                            \\ \cline{1-2} \cline{4-6} 
\begin{tabular}[c]{@{}l@{}}Heavy \\ Crude\end{tabular}      & 10880                                                                   & \multicolumn{1}{l}{}           &                                                                                         & \multicolumn{1}{l}{}                                                                 & \multicolumn{1}{l}{}                                                                 \\ \cline{1-2}
No.6 Fuel                                                   & 21597                                                                   & \multicolumn{1}{l}{}           &                                                                                         & \multicolumn{1}{l}{}                                                                 & \multicolumn{1}{l}{}                                                                 \\ \cline{1-2}
\end{tabular}
\end{table*}
 Table \ref{Cleanup cost of types of oils for spills} provides an analysis of cleanup costs for spills in both the US and non-US regions based on oil type. The study considered the cost of over 200 spill cases outside the US by primary cleanup methodology (Dagmar Schmidt Etkin Environmental Research Consulting Winchester, Massachusetts, USA).

Technological advancements have created several research opportunities, but the efficiency of methods for cleaning up oil spills has not kept up with the challenging problems. Current methods for cleaning up oil spills \cite{ghara2022new,karmelich2023advancing,dhaka2021review,singh2020environmental} are not very effective since waves in the ocean are dynamic, which significantly reduces the effectiveness of physical containment techniques like booms \cite{dhaka2021review}. Other traditional works \cite{kim2021simultaneous,li2021multi,wang2023cyber} often involve physical barriers or chemical dispersants, which can be invasive and potentially harmful to marine ecosystems.
In addition to having limited efficacy because of the constantly shifting oceanic circumstances, existing approaches pose the potential of secondary pollution and damage to marine ecosystems. To properly manage and mitigate oil spills, novel, non-invasive, and ecologically friendly methods are extremely necessary.

\begin{figure*}
    \includegraphics[width=0.9\textwidth]{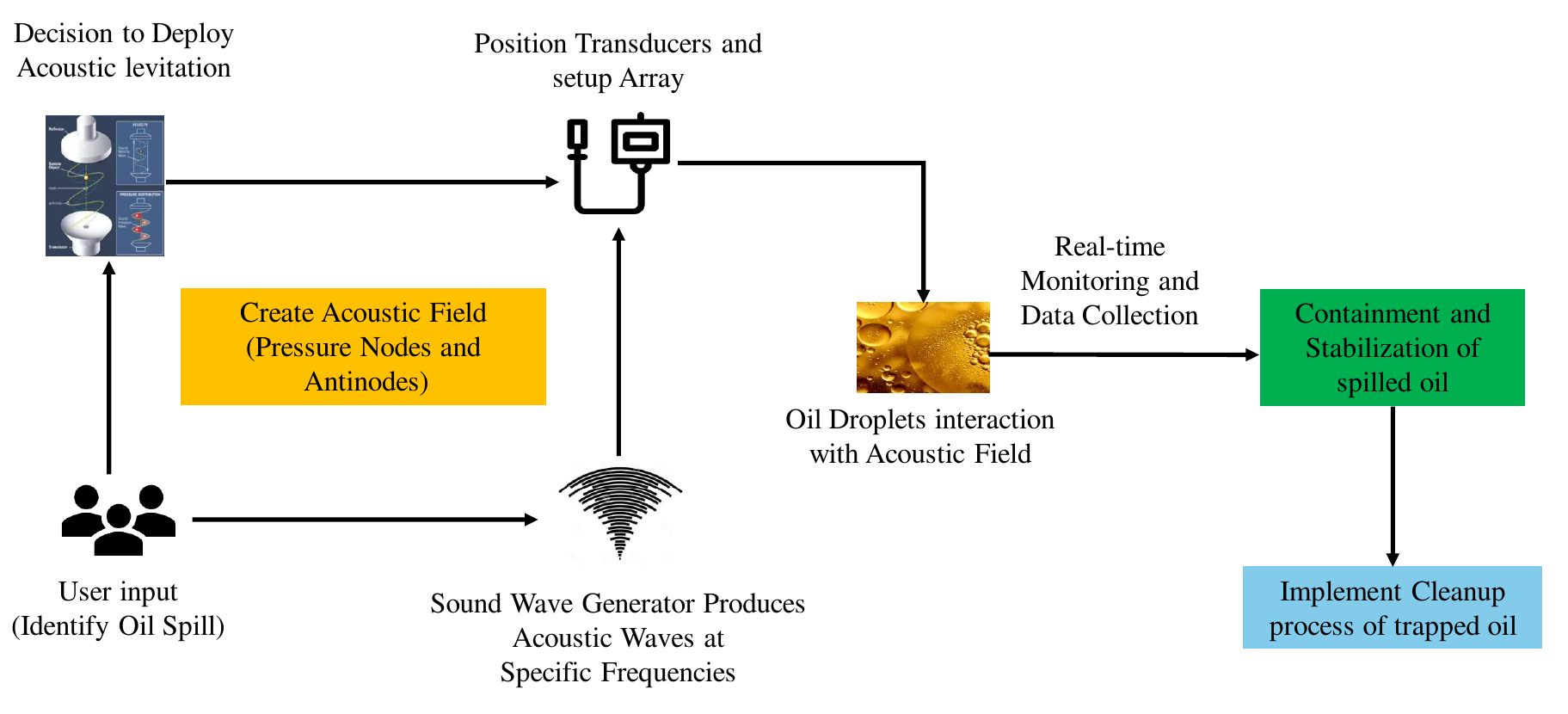}
    \caption{Conceptual Overview}
    \label{Conceptual Overview}
\end{figure*}

Acoustic levitation for environmental remediation proposed by our research represents a cutting-edge approach in addressing oil spills by harnessing the power of sound waves. It offers a non-contact method to manipulate and contain oil droplets on the water's surface. It generates acoustic waves that create pressure nodes and antinodes in the water \cite{gao2024stablelev,hawkes2022node}. Initially, a sound wave generator produces controlled acoustic waves, which are emitted into the water by an array of acoustic transducers strategically placed around the spill area. These waves generate regions of high and low acoustic pressure, termed nodes and antinodes respectively, on the water's surface. Oil droplets within these regions interact with varying pressures: they may be pushed toward nodes or stabilized within antinodes \cite{luo2018droplets}. This manipulation effectively confines and contains the oil droplets, preventing further dispersion. Oil droplets within this acoustic field can be effectively levitated or trapped at stable positions, preventing further spread across the water surface. This containment method is particularly advantageous because it minimizes direct contact with the oil and reduces the risk of secondary contamination or harm to marine life. Furthermore, it can aid in forecasting the behavior of oil spills. The proposed conceptual overview is depicted in Fig. \ref{Conceptual Overview}.

\subsection{Contribution in this research}

We aim to develop and implement effective, scalable, and environmentally safe methods to contain and forecast oil spills. In that pursuit, we make the following contributions:

\begin{itemize}
\item We introduce a pioneering innovative approach that leverages the unique properties of ultrasonic standing wave patterns to trap and contain spilled oil effectively.
\item  Our framework integrates real-time analysis with the on-field deployment of semi-submersible acoustic levitators, providing a novel strategy for oil spill management.

\end{itemize}
\section{Related Works}
 
 For responding to oil spills, mechanical techniques remain the primary approach \cite{ben2021recent,massaro2012optical}. The authors in \cite{massaro2012optical} proposed a method for detecting and quantifying oil spills in water that works by combining an optical fiber sensor with image processing. The accuracy of oil distribution characterization on the water surface is enhanced through the integration of energy-minimizing spline fitting and Hough transform algorithms.  Environmental factors like strong winds and choppy seas can severely limit the efficacy of mechanical procedures used in oil spill response, such as boom containment and skimming. These approaches can also be labor-intensive and laborious to cover huge spill regions, needing significant equipment and staff. Moreover, surface oil is the main focus of mechanical methods; subsurface and scattered oil remains untreated and might still be dangerous to the environment.

Owing to their capacity to absorb oil, synthetic sorbent materials have found extensive application in oil spill treatment \cite{hoang2021sorbent,kim2022role}. Hoang et al.
examined the crucial role sorbents play in cleaning up oil spills. It does so by highlighting new developments in sorbent-based oil-collecting equipment, like booms and skimmers, which have recovery efficiencies of up to 90\%. The work in \cite{kim2022role} covered developments in sorbent technology, and served as a resource for understanding practical applications that can be applied to significant oil spills and pinpointing unsolved issues. They presented the oil weathering process and oil sorption mechanisms along with their connection to the characteristics of the target pollutants. However, the absorbed oil turns utilized sorbent materials into hazardous trash. Further, they are less useful for large-scale spills since they have a limited absorption capacity and must be often replaced once saturated. Although natural sorbents \cite{zamparas2020application} are sustainable, their absorption capabilities are frequently lower than those of synthetic sorbents, requiring higher volumes for efficient cleanup. They may also be less resilient and more prone to deterioration, which lowers their overall effectiveness and longevity in attempts to clean up oil spills. To break down oil slicks into tiny droplets and promote natural biodegradation, chemical dispersants are frequently utilized in oil spill incidents \cite{adofo2022dispersants}. Their use is debatable, nevertheless, because of several serious environmental issues, prominent among which is their toxicity to marine life. Microorganisms are employed in bioremediation to break down oil contaminants in maritime settings \cite{arora2022bioremediation}. Even with its potential, the procedure is frequently inefficient and affected by several external variables- temperature, oxygen levels, salinity, pH levels, presence of Indigenous microbial communities, and hydrocarbon composition.
Although promising, nanotechnology-based technologies \cite{mishra2022emergence} for cleaning up oil spills can be expensive to develop and use on a big scale \cite{singh2020environmental,massaro2012optical}. 
 The use of nanotechnology in oil spill cleaning is reviewed in \cite{mishra2022emergence}, with an emphasis on how it can improve current techniques and solve environmental issues. Furthermore, there is a significant concern as it is yet unclear how the discharge of nanoparticles into marine ecosystems could impact human health and the environment. Table \ref{comparison} exhibits a comparison of the proposed methods with existing methods.

%
\begin{sidewaystable*}[!htbp]
\caption{Comparison of Proposed Method with State-of-Art Methods}
\label{comparison}
\begin{tabular}{|l|l|l|l|l|l|l|}
\hline
\multicolumn{1}{|c|}{\textbf{Techniques}} & \multicolumn{1}{c|}{\textbf{Types}}                                              & \multicolumn{1}{c|}{\textbf{Efficiency}}                                                                         & \multicolumn{1}{c|}{\textbf{Deployment}}                                              & \multicolumn{1}{c|}{\textbf{\begin{tabular}[c]{@{}c@{}}Environmental\\  Impact\end{tabular}}}            & \multicolumn{1}{c|}{\textbf{Cost}}                                                                             & \multicolumn{1}{c|}{\textbf{Usage Scenarios}}                                                         \\ \hline
\multirow{3}{*}{Booms}                    & \begin{tabular}[c]{@{}l@{}}Containment \\ Booms\end{tabular}                     & \multirow{3}{*}{\begin{tabular}[c]{@{}l@{}}High for Containment,\\ limited by wave action\end{tabular}}          & \multirow{3}{*}{\begin{tabular}[c]{@{}l@{}}Requires \\ anchoring\end{tabular}}        & \multirow{3}{*}{\begin{tabular}[c]{@{}l@{}}low,but can affect \\ marine life\end{tabular}}               & \multirow{3}{*}{\begin{tabular}[c]{@{}l@{}}Moderate, \\ varies\\ with boom \\ type/length\end{tabular}} & \multirow{3}{*}{\begin{tabular}[c]{@{}l@{}}Calm water and \\ shorelines\end{tabular}}                 \\ \cline{2-2}
                                          & \begin{tabular}[c]{@{}l@{}}Absorbent \\ Booms\end{tabular}                       &                                                                                                                  &                                                                                       &                                                                                                          &                                                                                                                &                                                                                                       \\ \cline{2-2}
                                          & Fire Booms                                                                       &                                                                                                                  &                                                                                       &                                                                                                          &                                                                                                                &                                                                                                       \\ \hline
\multirow{3}{*}{Skimmers}                 & \begin{tabular}[c]{@{}l@{}}Oleophilic \\ Skimmers\end{tabular}                   & \multirow{3}{*}{\begin{tabular}[c]{@{}l@{}}Limited by debris and \\ weather\end{tabular}}                        & \multirow{3}{*}{\begin{tabular}[c]{@{}l@{}}Deployed from \\ vessels\end{tabular}}     & \multirow{3}{*}{Low to Moderate}                                                                         & \multirow{3}{*}{\begin{tabular}[c]{@{}l@{}}High, requires \\ specialized \\ equipment\end{tabular}}            & \multirow{3}{*}{Best for calm sea}                                                                    \\ \cline{2-2}
                                          & \begin{tabular}[c]{@{}l@{}}Weir \\ Skimmers\end{tabular}                         &                                                                                                                  &                                                                                       &                                                                                                          &                                                                                                                &                                                                                                       \\ \cline{2-2}
                                          & \begin{tabular}[c]{@{}l@{}}Suction \\ Skimmers\end{tabular}                      &                                                                                                                  &                                                                                       &                                                                                                          &                                                                                                                &                                                                                                       \\ \hline
\multirow{2}{*}{Sorbents}                 & Natural                                                                          & \multirow{2}{*}{\begin{tabular}[c]{@{}l@{}}Effective for small \\ spills, less for large\end{tabular}} & \multirow{2}{*}{\begin{tabular}[c]{@{}l@{}}Mechanical or \\ manual\end{tabular}}      & \multirow{2}{*}{\begin{tabular}[c]{@{}l@{}}Depends on type\\ some create waste\end{tabular}} & \multirow{2}{*}{\begin{tabular}[c]{@{}l@{}}Low to \\ moderate\end{tabular}}   & \multirow{2}{*}{\begin{tabular}[c]{@{}l@{}}Best for small \\ localized spill\end{tabular}}            \\ \cline{2-2}
                                          & Synthetic                                                                        &                                                                                                                  &                                                                                       &                                                                                                          &                                                                                                                &                                                                                                       \\ \hline
\multirow{2}{*}{Dispersants}              & Chemical                                                                         & \multirow{2}{*}{\begin{tabular}[c]{@{}l@{}} May spread\\ contamination\end{tabular}}                              & \multirow{2}{*}{\begin{tabular}[c]{@{}l@{}}Sprayed from \\ aircraft\end{tabular}}     & \multirow{2}{*}{\begin{tabular}[c]{@{}l@{}}Toxic to marine life,\\ affects water quality\end{tabular}}   & \multirow{2}{*}{\begin{tabular}[c]{@{}l@{}}Moderate \\ to high\end{tabular}}                                   & \multirow{2}{*}{\begin{tabular}[c]{@{}l@{}}Suitable for large \\ spills\end{tabular}} \\ \cline{2-2}
                                          & Biological                                                                       &                                                                                                                  &                                                                                       &                                                                                                          &                                                                                                                &                                                                                                       \\ \hline
In-Situ Burning                           & \begin{tabular}[c]{@{}l@{}}Ignition with \\ Fire-resistant \\ Booms\end{tabular} & \begin{tabular}[c]{@{}l@{}}Effective in oil \\ reduction. Produce \\ smoke and residue\end{tabular}              & \begin{tabular}[c]{@{}l@{}}Ignition Source \\ and favorable \\ condition\end{tabular} & \begin{tabular}[c]{@{}l@{}}High, produces \\ smoke \\ and air pollution\end{tabular}                     & \begin{tabular}[c]{@{}l@{}}Moderate\\  to high\end{tabular}                                                    & \begin{tabular}[c]{@{}l@{}}Best for offshore\\ and calm \\ conditions\end{tabular}                    \\ \hline
Nano                                      & \begin{tabular}[c]{@{}l@{}}Nano \\ materials\end{tabular}                        & Potential for high                                                                                               & \begin{tabular}[c]{@{}l@{}}Utilizes \\ engineered \\ nanomaterials\end{tabular}       & Can minimize                                                                                             & \begin{tabular}[c]{@{}l@{}}Initial\\ deployment \\ cost is very\\  high\end{tabular}                           & \begin{tabular}[c]{@{}l@{}}Best suited for \\ targeted scenarios\end{tabular}                         \\ \hline
Current research                                  & \begin{tabular}[c]{@{}l@{}}Acoustic \\ Waves\end{tabular}                        & High efficiency                                                                                                  & \begin{tabular}[c]{@{}l@{}}Easy, deploy \\ transducers\end{tabular}                   & \begin{tabular}[c]{@{}l@{}}Minimal, \\ non-intrusive\\ method\end{tabular}                               & \begin{tabular}[c]{@{}l@{}}Lower \\ operational \\ cost\end{tabular}                                           & \begin{tabular}[c]{@{}l@{}}Moderately rough\\ seas. calm seas, \\ shorelines, offshore\end{tabular}   \\ \hline
\end{tabular}
\end{sidewaystable*}

\section{System Model}

\begin{sidewaysfigure*}[!htbp]
    \includegraphics[width=1\textwidth]{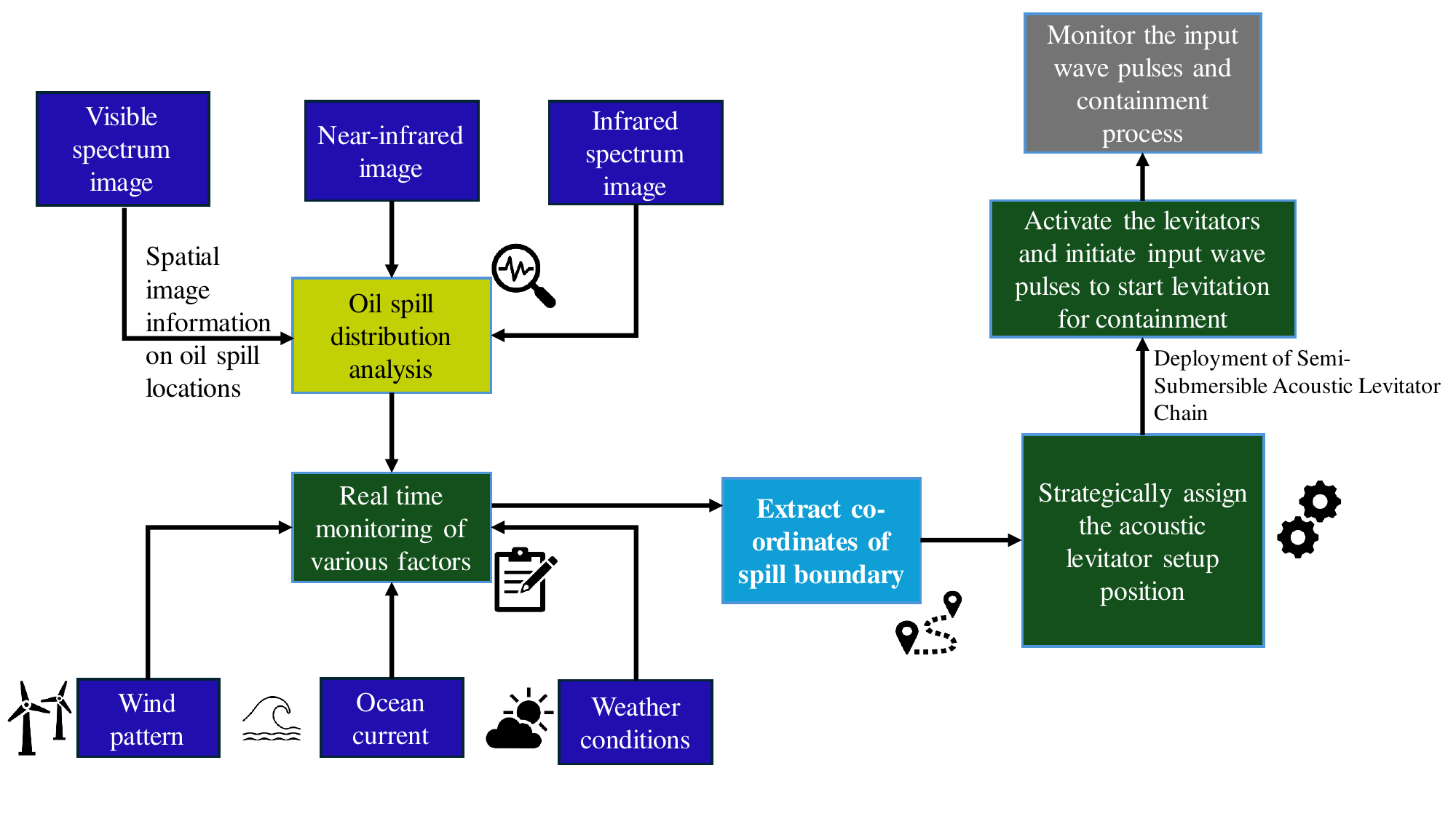}
    \caption{Flow Chart for the Proposed Methodology}
    \label{systemmodel}
\end{sidewaysfigure*}

\begin{figure}[ht]
\centering
    \includegraphics[width=0.6\textwidth, height=5cm]{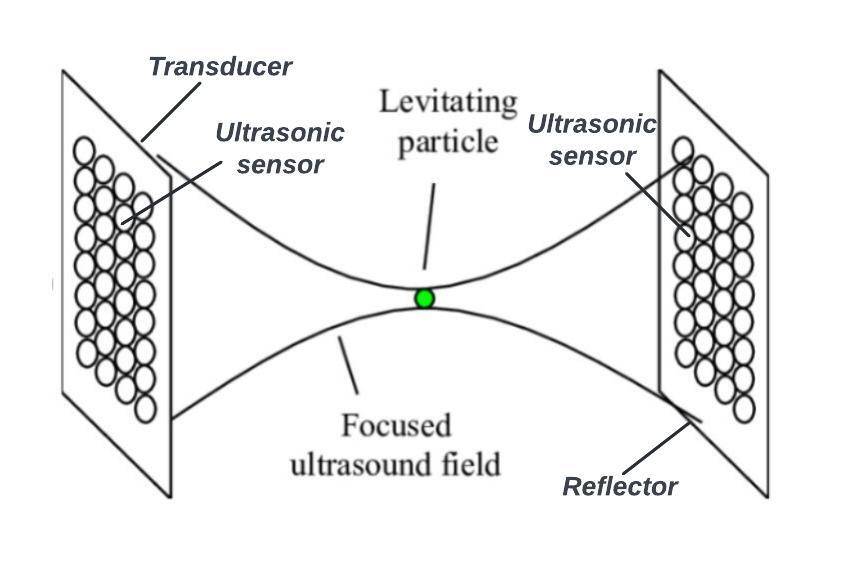}
    \caption{ Proposed Levitator Model}
    \label{proposed levitator model}
\end{figure}
The overall workflow of the proposed model depicted in Figure \ref{systemmodel}. It can be grouped into two components- Oil Spill Identification and Monitoring and Semi-Submersible Acoustic
Levitator chain. Obtaining input images from the visible, near-infrared, and infrared spectra is the first step in the procedure. The oil spill dispersion analysis system uses these images to determine the volume and direction of the spill. After conducting this analysis, a choice is made about using machine learning (ML) techniques for additional refinement. A real-time monitoring system receives this data and uses it to continuously track the oil spill progression. The system goes into the containment phase after the oil spill has been identified and tracked. Each levitator setup is moved to its appropriate spot using individual controls once the positions have been assigned. The levitation process is started by turning on the levitators and triggering input wave pulses once they are in position. The architecture's adaptability allows scalability from small-scale incidents to large disasters, offering a promising solution to enhance containment efficiency and minimize ecological impact compared to traditional cleanup methods. Figure \ref{proposed levitator model} illustrates the proposed levitator model.

\subsection{Oil Spill Identification and Monitoring:} The precise location, in-the-moment oil spill zone analysis, and spill expansion forecasting are performed by this component. The process entails utilizing multispectral imagery and data processing technologies to determine the amount of the spill and undertake dynamic analysis of the persistent oil leak. Acquiring satellite or aerial imaging to detect oil spills in a particular region involves several steps:
\begin{itemize}
    \item Identify the Region of Interest (ROI): Ascertain the region in which oil spills are most likely or anticipated to occur. This depends on factors like being close to industrial operations, offshore drilling sites, shipping lanes, or areas with known environmental hazards.
    \item Select Satellite or Aerial Platform: To find and track oil spills in the ROI, select a satellite or aerial platform that satisfies the requirements for both spatial and spectral resolution. This considers factors such as data availability, coverage region, revisit frequency, and sensor capabilities.
    \item Acquire Imagery: Initiate picture acquisition over the ROI by getting in touch with satellite imaging providers or aerial survey services. Indicate the ideal acquisition conditions, including the time and date of the image, the spectral bands, the spatial resolution, and any restrictions on cloud cover.
    \item Image Processing and Analysis: Preprocess and examine the imagery after it has been obtained to find any possible oil spills. To identify oil droplets from other surface features, apply image processing techniques such as spectral unmixing, supervised or unsupervised categorization, and change detection algorithms.
    \item Validation and Verification: Verify the identified oil spills visually or by having professionals confirm them using ground truth data, if feasible. Examine the precision and dependability of the location and magnitude of the identified oil spills.
    \item Data Integration and Visualization: Integrate the oil spill detection results with other relevant geospatial data, such as environmental conditions, ocean currents, wind patterns, and shipping traffic. Visualize the detected oil spills on maps or spatial databases for further analysis and decision-making.
    \item Monitoring and Follow-Up: Continuously monitor the region for new oil spills and changes in existing spills using satellite or aerial imaging. Monitoring system continuously receives data from the multispectral imaging system and data analytical tools, providing up-to-date information on the oil spill location and size. 
    \item Multispectral Imaging System: Multispectral imaging involves capturing images at multiple wavelengths across the electromagnetic spectrum. These images can reveal information about the composition and characteristics of the surface being observed. By using sensors that capture images in different spectral bands (visible, near-infrared, infrared, etc.), the multispectral imaging system can detect the presence of oil on the water's surface. The unique spectral signature of oil allows it to be distinguished from water and other materials. 
    \item Data Analytical Tools: Data analytical tools process the information collected by the multispectral imaging system to calculate the extent of the spill and analyze the ongoing oil leakage. These tools employ algorithms and models to analyze the multispectral images in real-time. They identify and segment areas of the image corresponding to the oil spill, calculate its area and volume, and track its movement over time. Advanced machine learning algorithms can improve the accuracy of spill detection and analysis by continuously learning from new data.
    \item Forecasting Models: Forecasting models utilize historical data and real-time observations to predict the future expansion of the oil spill. These models analyze factors such as wind patterns, ocean currents, weather conditions, and the characteristics of the spilled oil to forecast its trajectory and spread. By simulating different scenarios and incorporating uncertainty estimates, they provide decision-makers with probabilistic forecasts of where the oil spill is likely to move in the coming hours or days.
\end{itemize}
\subsubsection{Metrics for Real-Time Analysis}
Metrics to be taken into account for deriving the comprehensive formula for the real-time oil spill rate concerning wind speed and oil type include viscosity, density, and features of the water body. Oil spreads differently across water depending on wind speed \cite{ju2022mathematical, andrade2020acoustic} i.e 
 $ Oil Spread \propto W $, W is the wind speed.
The viscosity of the oil and the density difference between oil and water particles create buoyancy forces i.e  $ Oil Spread \propto \frac{\eta}{\rho oil} - \frac{1}{\rho water}$, $\eta$ is the oil viscosity, \(\rho\)water is the water density, and \(\rho\)oil is the oil density. The oil spill rate (OSR) can be determined using the equation \ref{eq1} \cite{andrade2016acoustic,ji2020influence}, where A is the spreading constant.
\begin{equation} \label{eq1}
OilSpillRate(OSR)= A W \left(\frac{\eta}{\rho oil} - \frac{1}{\rho water} \right)                    \end{equation}

Acoustic Radiation Pressure (ARP) can be calculated using the equation \ref{eq2} \cite{sarvazyan2021acoustic}, where I  is the intensity of the sound wave, and c  is the speed of sound in the medium.
\begin{equation} \label{eq2}
Acoustic Radiation Pressure(ARP) = \frac{2I}{c}     \end{equation}

Acoustic Radiation Force (ARF), produced when the medium receives momentum from the ultrasonic wave can be calculated using the equation \ref{eq3} \cite{andrade2016acoustic}, where \(\rho\)o is the density of oil, \(\rho\)w is the density of water, g is the acceleration due to gravity, and r is the radius of the oil droplet.
\begin{equation} \label{eq3}
Acoustic Radiation Force(ARF) = \frac{4 \pi r^3 g}{3}   ( \rho o- \rho w)  \end{equation}
\subsection{On-Field Deployment of Semi-Submersible Acoustic Levitator chain:} It is a chain of semi-submersible acoustic levitators (composed of Transducer Array, Reflector Plate, Node Generator, Floatation Devices, Frequency Generator) and feedback-sensors strategically positioned on the plotted coordinates obtained from the above component to form an effective barrier around the oil spill. These levitators create ultrasonic standing wave patterns in the water, effectively trapping the spilled oil and leaving water since the maintained frequency can hold up the lower denser oil droplets within the nodes of the standing wave. This acts as a containment barrier, preventing further spreading of the oil.  
\begin{itemize}
    \item Transducer Array: The transducer array consists of multiple piezoelectric transducers arranged in a specific configuration. When electrical energy is applied, these transducers generate ultrasonic waves in the water. Figure \ref{transducer} illustrates a schematic representation of the array.
    \begin{figure}
    \centering
    \includegraphics[width=0.4\textwidth]{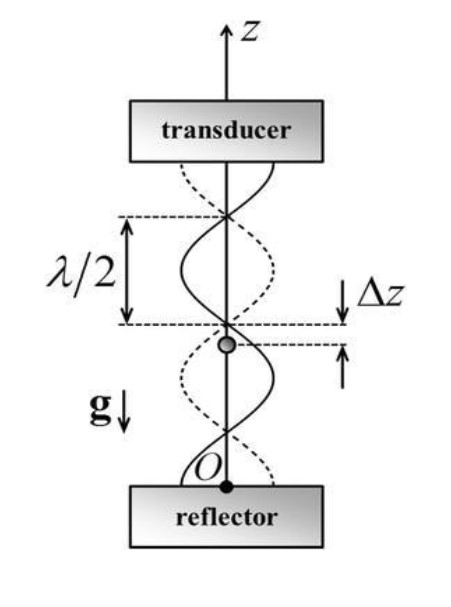}
    \caption{Schematic Representation of Transducer Array}
    \label{transducer}
\end{figure}
\item Reflector Plate: It directs and focuses ultrasonic waves towards the desired area and is positioned opposite to the transducer. It reflects the ultrasonic waves generated by the transducers towards the surface of the water, enhancing their effectiveness in creating standing wave patterns.
\item Node Generator: It creates standing wave patterns with nodes and antinodes by adjusting the phase and amplitude of ultrasonic waves from the transducer array. Nodes are regions of minimal displacement where denser oil particles accumulate, while antinodes are regions of maximum displacement.
\item Floatation Devices: It maintains the acoustic levitator at the desired depth in the water. Buoyant materials or devices are attached to the acoustic levitator to ensure that it remains at the appropriate depth in the water column. This helps to optimize the effectiveness of the standing wave patterns in trapping oil particles.
\item Frequency Generator: It regulates the transducer array's ultrasonic wave emission frequency. The transducer array receives electrical signals from the frequency generator, which modifies them to precisely control the ultrasonic frequency. To produce standing wave patterns with the ideal node spacing for effective oil particle trapping, frequency optimization is imperative.
\item Feedback Sensors: Provide real-time feedback on environmental conditions and acoustic wave properties. Sensors such as pressure sensors, temperature sensors, and hydrophones are integrated into the acoustic levitator to monitor water depth, temperature, and acoustic wave characteristics. This feedback information helps optimize ultrasonic frequency and standing wave patterns for maximum effectiveness.

(i) Pressure Sensors:
Monitor water pressure at various depths to ensure that the acoustic levitator remains at the desired depth for optimal performance. Pressure sensors can be placed at different points along the acoustic levitator chain to measure water pressure.

(ii) Temperature Sensors:
Monitors water temperature to assess environmental conditions that may affect the effectiveness of the acoustic levitator. Temperature sensors measure water temperature in the vicinity of the acoustic levitator. Temperature variations can impact the density and viscosity of the water, affecting the propagation of ultrasonic waves and the formation of standing wave patterns.

(iii) Hydrophones:
Capture and analyze acoustic signals emitted by the transducer array to assess the quality and characteristics of standing wave patterns. Hydrophones are underwater microphones capable of detecting and recording ultrasonic waves generated by the acoustic levitator. They provide real-time feedback on the frequency, amplitude, and phase of the acoustic waves.

(iv) Dissolved Oxygen Sensors: Monitor dissolved oxygen levels in the water to assess the impact of the oil spill on aquatic life and ecosystem health. It measures the concentration of oxygen dissolved in water. Decrease in dissolved oxygen levels could indicate the presence of oil contaminants.

(v) Oil Content Analyzers: Detects and quantifies the concentration of oil particles suspended in the water to assess the effectiveness of containment and cleanup efforts. Oil content analyzers utilize optical or chemical methods to measure the concentration of oil droplets or particles in water samples.

\end{itemize}

\section{Proof Of Concept (POC) and Discussions}

 \begin{figure*}[!ht]
 \centering
    \includegraphics[width=1\textwidth]{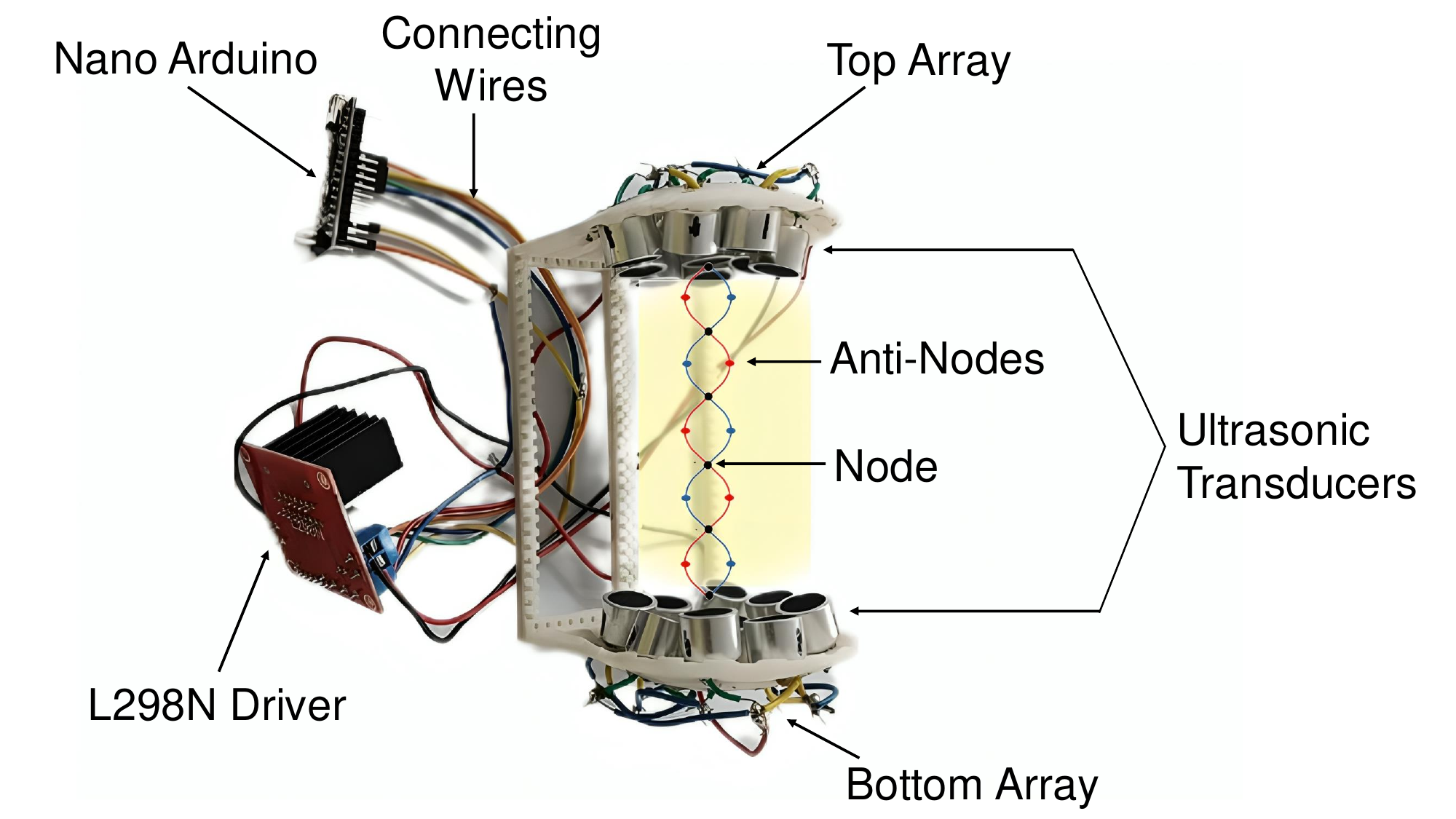}
    \caption{Experimental Hardware Setup for Evaluating the Proposed System}
    \label{levitator imge}
\end{figure*}
The experimental setup is illustrated in Figure \ref{levitator imge}. A rudimentary acoustic levitator was built with a range of 5V - 40KHz piezoelectric transducers. This device comprises two main components: a transducer that generates sound vibrations and a reflector, typically featuring concave surfaces to focus the sound waves. Our experimentation demonstrated the generation of standing waves and interference when the transducer's sound waves bounced off the reflector. We assessed the model's ability to produce enough ultrasonic force to suspend particles with a density of 20 $kg/m^3$. Then we evaluated how well it trapped 700 $kg/m^3$ of oil droplets. The scaled-down version of the levitator model effectively raised a 20 $kg/m^3$ foam particle following multiple trials. Subsequently, oil droplets were employed to evaluate its efficacy; Table 3 demonstrates the results of the investigation. The configuration employed for experimentation includes- number of transducers: 14, power output per transducer: 1 Watt, total power output: 14 Watts, Area selected for experiments: 0.1 $m^2$, wind speed: 343 m/s. The ARP for particle entrapment has been determined as:

\begin{align}
I &= \frac{14}{0.1} = 140 \text{ W/$m^2$} \nonumber \\
ARP &= \frac{2*140}{343} \approx 0.815 \text{Pa} \nonumber
\end{align}

We have taken into account the buoyancy and acoustic forces to calculate the necessary acoustic pressure to trap oil droplets with a density of 700 kg/m³ in water. Due to its lower density than water, oil will naturally rise to the top. Acoustic levitation can assist in capturing and directing these droplets throughout the water column. The formula for the acoustic radiation force can be used to estimate the necessary pressure, assuming that the oil droplets must be suspended against gravity and other forces. For a small oil droplet with a radius of 1 mm ( r = 0.001 m ), the ARF is obtained as:
\begin{align}
ARF &= \frac{4 \pi (0.001)^3 * 9.81}{3} (1000-700) \approx 1.23 * 10^{-5}   \text{ N}  \nonumber 
\end{align}
Considering that the droplet is in the standing wave node, the associated ARP required to oppose this force is calculated as :

\begin{align}
ARP &= \frac{ARF}{A} = \frac{1.23 * 10^{-5}}{\pi (0.001)^2}  \approx 3.91   \text{ Pa} \nonumber 
\end{align}
We experimented at low, medium, and high pressures, using this pressure value as the threshold. The results obtained are shown in the Table \ref{trial}.

\begin{table}[ht]
\centering
\caption{Experimental Values depicting the relationship between the Intensity of the Ultrasonic Pressure and the Proportion of Oil Droplets Trapped.
}
\label{trial}
\begin{tabular}{|c|l|c|c|c|}
\hline
\textbf{Trial} & \multicolumn{1}{c|}{\textbf{\begin{tabular}[c]{@{}c@{}}Pressure \\ Value\end{tabular}}} & \textbf{\begin{tabular}[c]{@{}c@{}}Initial Oil \\ Concen. \\ Trapped (\%)\end{tabular}} & \textbf{\begin{tabular}[c]{@{}c@{}}Final Oil \\ Concen. \\ Trapped (\%)\end{tabular}} & \textbf{\begin{tabular}[c]{@{}c@{}}Duration\\ (m)\end{tabular}} \\ \hline
1              & No                                                                                      & 0                                                                                       & 0                                                                                     & 30                                                              \\ \hline
2              & Low                                                                                     & 0                                                                                       & 7                                                                                     & 20                                                              \\ \hline
3              & Medium                                                                                  & 7                                                                                       & 10                                                                                    & 10                                                              \\ \hline
4              & High                                                                                    & 10                                                                                      & 11.5                                                                                  & 10                                                              \\ \hline
\end{tabular}
\end{table}

\subsection{Inferences}

The POC demonstrates that acoustic levitation can trap particles with densities lower than water, such as oil droplets. However, achieving the necessary acoustic radiation pressure to contain oil spills requires optimization of the power output, frequency, and configuration of the levitator array. This can be achieved by:
\begin{itemize}
    \item Increasing Power Output: By increasing the number of transducers or the power per transducer, the total power output can be increased. For instance, doubling the number of transducers to 28 would double the intensity and pressure.
    \item Optimizing Frequency: The frequency of the transducers affects the wavelength and the formation of standing waves. Ultrasonic frequencies in the range of 40-60 kHz are typically used for levitation. Fine-tuning the frequency can help achieve more efficient trapping.
\item Adjusting Area: Reducing the area over which the sound waves are focused can increase the intensity, thereby increasing the acoustic radiation pressure.
\item Use of Reflectors: More efficient reflectors can enhance the formation of standing waves, leading to higher pressure zones.

\end{itemize}

Further research and development are necessary to scale this technology for practical oil spill containment applications.

 \section{Conclusion}

The potential of ultrasonic levitation as a useful strategy for oil spill containment and forecasting is demonstrated by this research. Acoustic levitation is a reliable and ecological alternative to conventional oil spill cleanup techniques, as it employs sound waves to effectively detach oil droplets from water. The method's viability was confirmed by the proof-of-concept trials, which revealed a noteworthy correlation between the strength of ultrasonic pressure and the oil droplet collection efficiency. The non-invasiveness of this approach and its capacity to function in regulated settings make it unique. Even so, to optimize the technology for realistic, large-scale applications, further research and development needs to be conducted.

\section{Acknowledgments}

The authors would like to thank the editors and reviewers. The authors would like to thank Kota Reddy, Vice Chancellor, VIT-AP University, Jagadish Chandra Mudiganti, Registrar, VIT-AP University and Hari Seetha, Director, Centre of Excellence, Artificial Intelligence and Robotics (AIR)  for their support. Special thanks to the team members of the Artificial and Robotics (AIR) Center, VIT-AP University.


\bibliographystyle{elsarticle-num} 
\bibliography{Ocean_Oil_Splil_Seggregation}

\begin{thebibliography}{10}
\expandafter\ifx\csname url\endcsname\relax
  \def\url#1{\texttt{#1}}\fi
\expandafter\ifx\csname urlprefix\endcsname\relax\def\urlprefix{URL }\fi
\expandafter\ifx\csname href\endcsname\relax
  \def\href#1#2{#2} \def\path#1{#1}\fi

\bibitem{news_2024}
{ITOPF}, Oil tanker spill statistics 2023,
  {https://www.itopf.org/news-events/news/oil-tanker-spill-statistics-2023/},
  accessed on 2024-2-24 (Jan. 2024).

\bibitem{de2021immediate}
M.~de~Oliveira~Estevo, P.~F. Lopes, J.~G.~C. de~Oliveira~J{\'u}nior, A.~B.
  Junqueira, A.~P. de~Oliveira~Santos, J.~A. da~Silva~Lima, A.~C.~M. Malhado,
  R.~J. Ladle, J.~V. Campos-Silva, Immediate social and economic impacts of a
  major oil spill on brazilian coastal fishing communities, Marine Pollution
  Bulletin 164 (2021) 111984.

\bibitem{ghara2022new}
F.~M. Ghara, S.~B. Shokouhi, G.~Akbarizadeh, A new technique for segmentation
  of the oil spills from synthetic-aperture radar images using convolutional
  neural network, IEEE Journal of Selected Topics in Applied Earth Observations
  and Remote Sensing 15 (2022) 8834--8844.

\bibitem{karmelich2023advancing}
C.~Karmelich, Z.~Wan, W.~Tian, E.~Crooke, X.~Qi, A.~Carroll, K.~Konstas,
  C.~Wood, Advancing hyper-crosslinked materials with high efficiency and
  reusability for oil spill response, Scientific Reports 13~(1) (2023) 9779.

\bibitem{dhaka2021review}
A.~Dhaka, P.~Chattopadhyay, A review on physical remediation techniques for
  treatment of marine oil spills, Journal of Environmental Management 288
  (2021) 112428.

\bibitem{singh2020environmental}
H.~Singh, N.~Bhardwaj, S.~K. Arya, M.~Khatri, Environmental impacts of oil
  spills and their remediation by magnetic nanomaterials, Environmental
  nanotechnology, monitoring \& management 14 (2020) 100305.

\bibitem{kim2021simultaneous}
D.-h. Kim, R.~Mauchauff{\'e}, J.~Kim, S.~Y. Moon, Simultaneous, efficient and
  continuous oil--water separation via antagonistically functionalized
  membranes prepared by atmospheric-pressure cold plasma, Scientific Reports
  11~(1) (2021) 3169.

\bibitem{li2021multi}
B.~Li, H.~Ma, B.~Mao, Multi-user online three-dimensional marine oil spill
  crisis response system, in: IEEE/WIC/ACM International Conference on Web
  Intelligence and Intelligent Agent Technology, 2021, pp. 17--21.

\bibitem{wang2023cyber}
Y.~Wang, X.~Chen, L.~Wang, Cyber-physical oil spill monitoring and detection
  for offshore petroleum risk management service, Scientific Reports 13~(1)
  (2023) 4586.

\bibitem{gao2024stablelev}
L.~Gao, G.~Christopoulos, P.~Mittal, R.~Hirayama, S.~Subramanian, Stablelev:
  Data-driven stability enhancement for multi-particle acoustic levitation, in:
  Proceedings of the CHI Conference on Human Factors in Computing Systems,
  2024, pp. 1--11.

\bibitem{hawkes2022node}
J.~J. Hawkes, S.~Maramizonouz, C.~Jia, M.~Rahmati, T.~Zheng, M.~B. McDonnell,
  Y.-Q. Fu, Node formation mechanisms in acoustofluidic capillary bridges,
  Ultrasonics 121 (2022) 106690.

\bibitem{luo2018droplets}
X.~Luo, J.~Cao, H.~Yin, H.~Yan, L.~He, Droplets banding characteristics of
  water-in-oil emulsion under ultrasonic standing waves, Ultrasonics
  sonochemistry 41 (2018) 319--326.

\bibitem{ben2021recent}
S.~ben Hammouda, Z.~Chen, C.~An, K.~Lee, Recent advances in developing
  cellulosic sorbent materials for oil spill cleanup: A state-of-the-art
  review, Journal of Cleaner Production 311 (2021) 127630.

\bibitem{massaro2012optical}
A.~Massaro, A.~Lay-Ekuakille, D.~Caratelli, I.~Palamara, F.~C. Morabito,
  Optical performance evaluation of oil spill detection methods: Thickness and
  extent, IEEE transactions on instrumentation and measurement 61~(12) (2012)
  3332--3339.

\bibitem{hoang2021sorbent}
A.~T. Hoang, X.~P. Nguyen, X.~Q. Duong, T.~T. Huynh, Sorbent-based devices for
  the removal of spilled oil from water: a review, Environmental Science and
  Pollution Research 28 (2021) 28876--28910.

\bibitem{kim2022role}
H.~Kim, G.~Zhang, T.~M. Chung, C.~Nam, A role for newly developed sorbents in
  remediating large-scale oil spills: Reviewing recent advances and beyond,
  Advanced Sustainable Systems 6~(1) (2022) 2100211.

\bibitem{zamparas2020application}
M.~Zamparas, D.~Tzivras, V.~Dracopoulos, T.~Ioannides, Application of sorbents
  for oil spill cleanup focusing on natural-based modified materials: A review,
  Molecules 25~(19) (2020) 4522.

\bibitem{adofo2022dispersants}
Y.~K. Adofo, E.~Nyankson, B.~Agyei-Tuffour, Dispersants as an oil spill
  clean-up technique in the marine environment: A review, Heliyon 8~(8) (2022).

\bibitem{arora2022bioremediation}
S.~Arora, S.~Saxena, D.~Sutaria, J.~Sethi, Bioremediation: An ecofriendly
  approach for the treatment of oil spills, in: Advances in Oil-Water
  Separation, Elsevier, 2022, pp. 353--373.

\bibitem{mishra2022emergence}
S.~Mishra, G.~Chauhan, S.~Verma, U.~Singh, The emergence of nanotechnology in
  mitigating petroleum oil spills, Marine Pollution Bulletin 178 (2022) 113609.

\bibitem{ju2022mathematical}
X.~Ju, Z.~Li, B.~Dong, X.~Meng, S.~Huang, Mathematical physics modelling and
  prediction of oil spill trajectory for a catenary anchor leg mooring (calm)
  system, Advances in Mathematical Physics 2022~(1) (2022) 3909552.

\bibitem{andrade2020acoustic}
M.~A. Andrade, A.~Marzo, J.~C. Adamowski, Acoustic levitation in mid-air:
  Recent advances, challenges, and future perspectives, Applied Physics Letters
  116~(25) (2020).

\bibitem{andrade2016acoustic}
M.~A. Andrade, A.~L. Bernassau, J.~C. Adamowski, Acoustic levitation of a large
  solid sphere, Applied Physics Letters 109~(4) (2016).

\bibitem{ji2020influence}
H.~Ji, M.~Xu, W.~Huang, K.~Yang, The influence of oil leaking rate and ocean
  current velocity on the migration and diffusion of underwater oil spill,
  Scientific Reports 10~(1) (2020) 9226.

\bibitem{sarvazyan2021acoustic}
A.~P. Sarvazyan, O.~V. Rudenko, M.~Fatemi, Acoustic radiation force: a review
  of four mechanisms for biomedical applications, IEEE Transactions on
  Ultrasonics, Ferroelectrics, and Frequency Control 68~(11) (2021) 3261--3269.

\end{thebibliography}

\end{document}